# A parsimonious structure model for the icosahedral quasicrystal $Cd_{5.7}Yb$

**M. Feuerbacher**[a]

[a] Ernst Ruska-Centre for Microscopy and Spectroscopy with Electrons ER-C-1, Forschungszentrum Jülich GmbH, 52425 Jülich, Germany

**Synopsis** The paper presents a parsimonious structure model for the icosahedral $Cd_{5.7}Yb$ quasicrystal. It employs only spherical and elliptical occupation domains, which significantly decreases the number of parameters and the required calculation effort. Still, the model describes the characteristic structural features with considerable accuracy.

**Abstract** A compact structure model for the icosahedral phase $Cd_{5.7}Yb$ is presented. The model is based on a higher-dimensional description and a cut-and-projection procedure to generate the atom distribution in three-dimensional physical space. It strictly employs only spherical or elliptical occupation domains, ensuring a low number of parameters and low calculation cost. In the presented form the model has seven adjustable parameters. Despite its minimalistic setting, the model reproduces available experimental data, such as chemical composition, mass density and electron density distribution, as well as the cluster structure in physical space very well. Nevertheless, the approximation of the occupation domains by spheres and ellipses evidently implies limitations in accuracy and universality. The model can be adapted for the description of other Tsai-type icosahedral phases.

**Keywords: Structure model; Quasicrystal; i-CdYb; Tsai phase.**

## 1. Introduction

Quasicrystals are nonperiodic long-range ordered solids possessing non-crystallographic symmetries (Shechtman et al., 1984). Icosahedral quasicrystals (i-QCs), e.g., are quasiperiodic in three dimensions and have six fivefold rotational axes that are arranged according to the symmetry of an icosahedron. The lack of translational symmetry impedes a finite structure description in three-dimensional (3D) physical space. A finite description is enabled by the well-established framework of higher-dimensional crystallography (Duneau and Katz, 1985), which for i-QCs makes use of a six-dimensional (6D) hyperspace. In hyperspace, the structure is defined through a periodic arrangement of 3D objects, the so-called occupation domains (ODs). The 6D space is composed by a direct sum of two 3D spaces – physical space, which represents our common physical world, and perpendicular space, which is habituated by the ODs. The atomic structure in physical space is obtained from this 6D description through a well-defined procedure (Kramer and Neri, 1984), known as “cut and projection”. In a rudimentary program code, this can be implemented by constructing trial positions in 6D space as arbitrary sums of its basis vectors. The 6D coordinates of all OD centers in the corresponding 6D unit cell are projected onto perpendicular space. If the perpendicular-space coordinate is smaller than the extension of the OD along the same direction, an actual atom position is generated at the intersection of the OD and physical space.

The ODs used for the construction of i-QCs are always 3D objects but can take different shapes. In the most generic case, the construction of a 3D Amman-Kramer-Neri tiling, there is only one OD, a rhombic triacontahedron, placed on the vertices of the 6D lattice (Katz and Duneau, 1986). This faceted polyhedral OD can be approximated by a sphere, which greatly simplifies the calculations involved in the cut-and-projection procedure; for the case of a sphere, only the radius has to be compared with the length of the perpendicular-space coordinate and no direction dependencies have to be taken into account. This approximation comes with some disadvantages though: since the symmetry of a spherical OD does not match its site symmetry, overlaps and gaps with neighboring ODs can occur, leading to the generation of unphysically close distances and vacant positions in physical space, respectively. This however holds only, if full spheres are used as ODs. Truncation of the ODs can be applied in order to eliminate overlaps with neighboring ODs and avoid the generation of short distances (De Boissieu et al., 1991). If for the definition of these truncations only spheres are employed, the calculation costs do not increase much. In a sense, truncation of spherical ODs can be considered a way to achieve a closer match of an OD to its site symmetry.

The ODs of real quasicrystals, which are always binary, ternary or even multi-elemental alloys, contain not only spatial but also chemical information. The latter is encoded by subdividing the ODs into different regions, which generate different species of atoms on different sites in physical space. For example, the ODs can be decomposed into concentric shells or polygons, in order to define distinct regions that can be allocated to specific elements.

A well-known example of a higher-dimensional model for a real quasicrystal was developed by Boudard et al. (1992) for i-AlPdMn. The model employs three spherical ODs located on high-symmetry positions of the 6D lattice, two of which are subdivided into concentric shells.

On the whole, the model reproduces the characteristic features of the physical-space structure very well, but since this it employs only spherical ODs it generates a certain density of short distances. It has nevertheless successfully served as a versatile workhorse model, that has been applied and modified for several purposes, e.g. for modelling of dislocations (Yang et al. 1998), surfaces (Ledieu et al., 2001) or the vibrational density of states (Cordelli and Gallo, 1995).

A similarly constructed model was established for i-ZnMgHo (Ishimasa et al., 2004). In this model edge-center positions in the 6D unit cell were populated, which possess a lower site symmetry. Therefore, it required the introduction of ellipsoidal ODs.

More elaborate models, employing faceted ODs were developed for i-AlPdMn (Quiquandon and Gratias, 2006, Yamamoto et al., 2003), which are more accurate, and capable of generating realistic mass densities while avoiding the generation of unphysically short distances.

An outstanding, more recent structure model was developed by Takakura et al. (2007) for i-CdYb. Its ODs are composed by a highly dedicated perpendicular-space arrangement of Henley polyhedra (Henley, 1986), maintaining all hyperlattice site symmetries. The individual Henley polyhedra constitute OD regions defining physical-space positions and atomic species at great detail. This leads to a model of *hitherto* unknown accuracy – the calculated chemical composition and mass density are in excellent agreement with the experimental values, and the physical-space structure displays all characteristic local features, in particular the Tsai clusters, which are structural building blocks with very specific concentric shell buildup. On top of the detailed perpendicular-space structure of the ODs, parts of the latter are subject to physical-space shifts. Such shifts of parts of the OD are permitted as long as they maintain the local hyperspace site symmetry, and allow for the very dedicated arrangement of physical-space atom positions (De Boissieu et al., 1990). The extreme accuracy of the model comes at a cost, however. The asymmetric parts of the ODs used in the model are constructed using numerous Henley polyhedra, each of which has 150 facets. Even with an effective algorithm, taking advantage of all symmetries, the calculation effort is elaborate and time consuming. Also, reproduction of the model in an own code, which may be required to adapt it to the individual researchers needs, is extremely demanding.

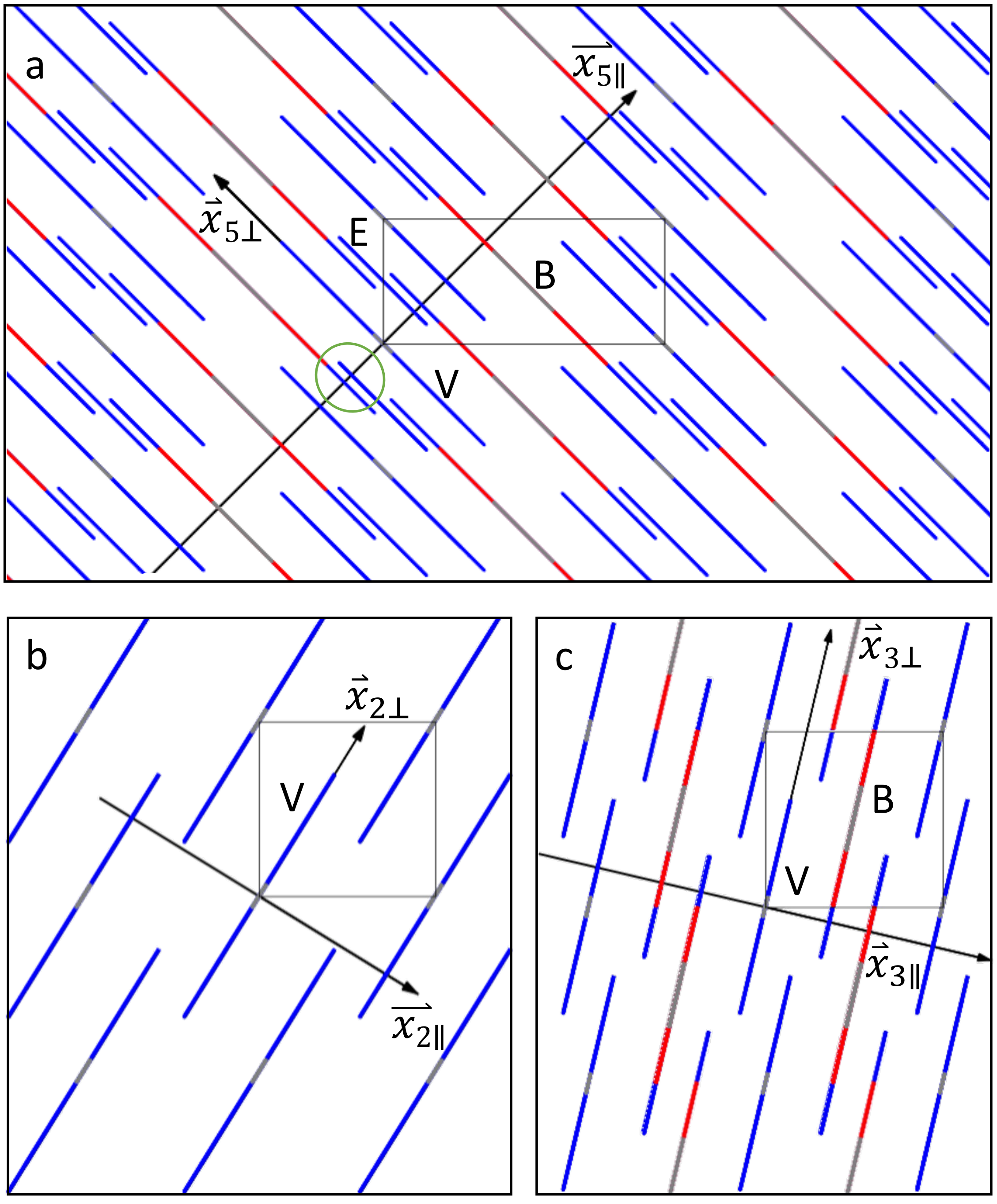

**Figure 1** Basic setting of the model displayed as traces of the spheres representing the ODs in fivefold (a), twofold (b) and threefold (c) rational sections of 6D space. These sections contain a pair of physical- and perpendicular-space vectors of the indicated symmetry. Instances of the ODs V, B, and E, forming the basis of the 6D unit cell, the projected outline of which is indicated, are marked. OD regions shown in blue and red generate Cd and Yb atoms, respectively. The hollow OD centers are shown in grey. The use of full spherical ODs leads to the creation of unphysical short distances in physical space (encircled in green).

With the present paper, I introduce an alternative model for i-CdYb, which is parsimonious in the sense that it operates with a strongly reduced number of parameters by using spherical and elliptical ODs only. Without exception, all model parameters are provided and the complete workings of the algorithm are described in full detail, so the functionality is unconditionally transparent and reproducible. A source code (in python) will be provided (Feuerbacher, 2025), but if necessary, reprogramming is straightforward. It is not the intention of this paper to provide a competing model to that of Takakura et al. (2007), and it is obvious that it can neither keep up with the latter's accuracy nor lessen its relevance. Nevertheless, I believe the availability of a simple, easy to use and versatile model, is very helpful for many occasions in everyday work. In this sense the model should be understood as a complementary *Ansatz* for cases, where utmost precision and fidelity is not required. Obvious potential applications of the model are fast physical-property calculations for cases where only basic structural characteristics are relevant. Also, the model will be beneficial for diffraction calculations, as it allows the calculation of essential features of diffraction patterns without the need to compute Fourier transforms of complex OD shapes and arrangements. Finally, the hands-on availability of a model that allows manageable manipulations in 6D hyperspace is required for modelling in cases where components in perpendicular space are critically relevant. This is e.g. the case for dislocations. It has been demonstrated (Levine et al., 1985) that dislocations in icosahedral quasicrystals are 6D objects, which possess strain-field components and hence Burgers-vector components in both physical and perpendicular space. Dislocation structure modelling in icosahedral quasicrystals imperatively involves a construction process, such as the removal or insertion of a hyperspace plane and strain-field definition in 6D, with subsequent cut-and-projection of the entire hyperspace object into physical space (Yang et al., 1998). Even though the on-demand availability of physical-space atom coordinates is generously and straightforwardly assured (Takakura 2025), this is not effectual in situations that require manipulation of the model in hyperspace.

Note that a model for i-Cd-Yb even simpler than the one presented in this paper has been previously used for the calculation of diffraction patterns (Uryu et al., 2024). No further specifications of the model or its performance in the generation of physical-space structure were provided, however.

## 2. Structure model

The design of the model is strongly oriented at the highly accurate Takakura model (Takakura et al., 2007), however under the condition that only spherical ODs or, if the site structure calls for it, ellipsoidal ODs are used. A cubic primitive hyperlattice with space group $Pm\overline{35}$ and a 6D lattice constant $a_0$ = 8.045 Å are assumed. ODs are placed at three high-symmetry positions in the 6D unit cell: on the corner position at (0,0,0,0,0,0), on the body-centered position ½(1,1,1,1,1,1) and on the edge-center position at ½(1,0,0,0,0,0), referred to as V, B and E, respectively. Due to the lower site symmetry of the edge-center positions, the E OD has a multiplicity of 6, i.e. individual instances at ½(1,0,0,0,0,0), ½(0,1,0,0,0,0), etc. have to be taken into account. The basic setting of the 6D unit cell is depicted in Fig. 1, not taking into account the truncations or physical-space shifts of the ODs that will be described below. The ODs are shown in rational sections of the 6D lattice, i.e. in planes which contain a pair of corresponding physical- and perpendicular-space vectors. In these planes, the spheres representing the ODs are then seen as traces with an extension only along the perpendicular-space direction. Regions of the ODs shown in blue and red generate Cd and Yb atoms, respectively. The hollow OD centers are shown in grey. In Fig. 1(a), a fivefold rational section spanned by the 6D vectors [0 1 1 1 1 $\bar{1}$] and [1 0 0 0 0 0], containing the fivefold vectors [0/1 -1/0 0/0] in physical and [1/0 0/1 0/0] in perpendicular space, is shown (using the notation of Cahn et al. (1986) for physical- and perpendicular-space coordinates). This plane contains traces of all three ODs. In (b) a twofold rational section spanned by [0 1 0 0 1 0] and [$\bar{1}$ 0 0 1 0 0], containing the twofold vectors [0/2 0/0 0/0] and [2/0 0/0 0/0] in physical and perpendicular space, respectively, is shown. Finally, (c) shows a threefold rational section spanned by [0 0 $\bar{1}$ 0 1 $\bar{1}$] and [$\bar{1}$ 1 0 1 0 0], containing the threefold vectors [1/0 0/0 $\bar{2}$/1] in physical and [0/1 0/0 $\bar{1}$/$\bar{2}$] in perpendicular space.

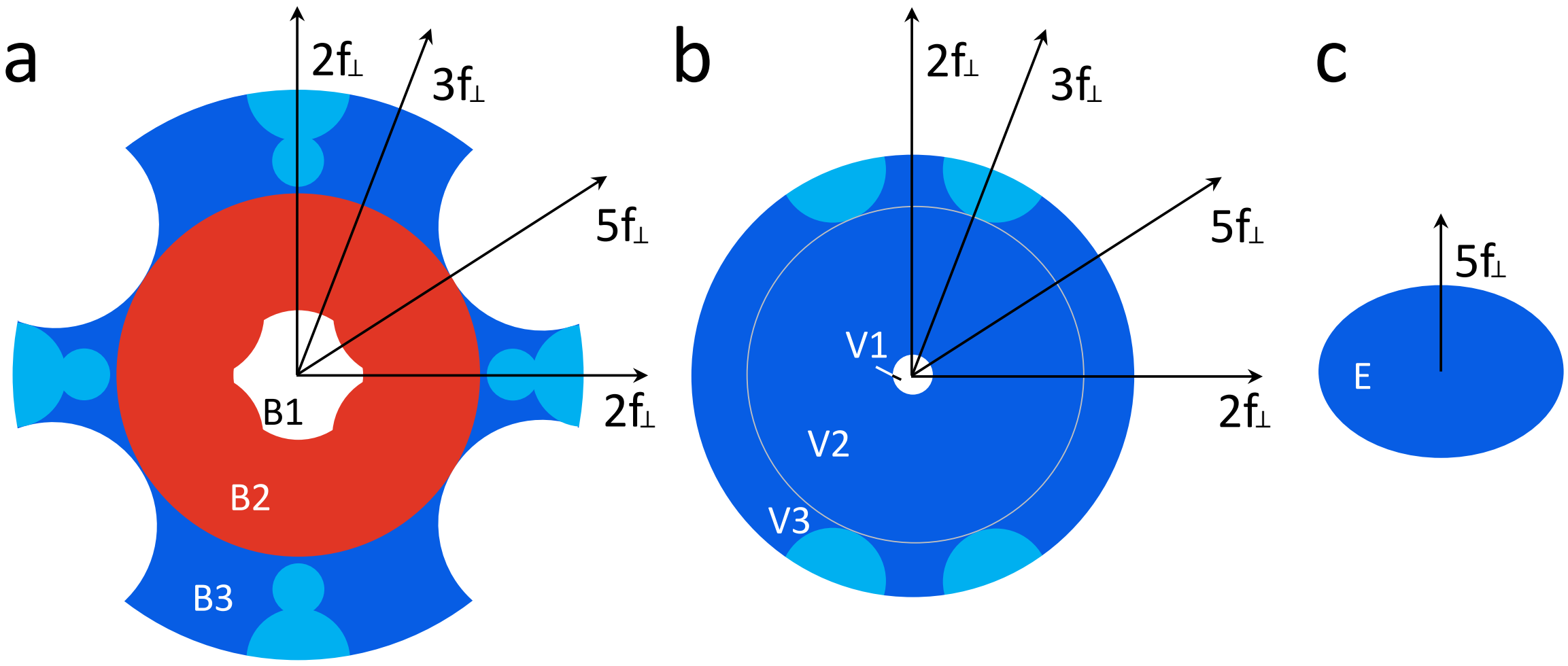

**Figure 2** Cross sections of the ODs of the model perpendicular to a twofold perpendicular-space direction. (a): B OD, consisting of a hollow center B1, a second shell B2 generating Yb positions (red) and a third shell B3 generating Cd positions (blue). The center and the third shell are truncated along the fivefold ($5f_{\perp}$) directions. (b): V OD, consisting of a hollow center V1 plus a second and third shell, V2 and V3, generating Cd positions (blue). The outer shells B3 and V3 include regions (light blue) along the twofold ($2f_{\perp}$) and threefold ($3f_{\perp}$) directions, respectively, generating positions subjected to physical-space shifts (see text). (c): E OD with ellipsoidal shape generating Cd positions. The short axis is parallel to a fivefold direction.

In Fig. 1, representing the basic setting of the model, the limitations of full spherical ODs are directly visible. As an example, consider the green encircled region: upon projection, the extension of the B and E ODs here lead to a physical-space distance of 0.67 Å between two Cd atoms, which is unphysically short. Correspondingly, the outer shell of the B OD employed in the model will be truncated.

Most formative for the physical-space structure is the B OD at the body-centered position, which is also the only OD that generates Yb atoms. This OD was constructed as a hollow inner sphere B1 of radius $rB_1$ surrounded by two further concentric shells B2 and B3 (Fig. 2a). The radius of the hollow center roughly corresponds to that of a Henley polyhedron, i.e. 3.0 Å or 0.372 $a_0$. The second shell of radius $rB_2$ leads to the generation of Yb atoms, the third shell of radius $rB_3$ leads to the generation of Cd atoms.

The OD on the B position employed in the Takakura model is shorter along the fivefold directions in perpendicular space, to an extent that there exist no outer Cd regions along these directions. In the present spherical model similar conditions are established by truncation: half spheres of radius $rB_3 - rB_2$ centered on the tangent of the OD at $rB_3$ are cut out of the outer shell in all fivefold directions corresponding to an icosahedron in perpendicular space. In perpendicular space, the B OD is thus a sphere with pronounced spherical concave dents along the fivefold directions, which at their deepest point touch the second shell B2. As such it approximates the outer form of the B OD displayed in Fig. 3b of Takakura et al. (2007). This truncation is crucial to avoid short distances along fivefold directions in the physical-space structure.

In a similar manner the hollow center is shaped: along the twofold directions the hollow center requires a certain extension in order to generate the distribution of empty Tsai-cluster centers in the fivefold physical-space planes. Along the fivefold directions, however, the same extension of the hollow center would lead to empty sites that are not consistent with the experimental electron density (Takakura et al., 2002). In the Takakura model, the center of the B OD is represented by the 'archetype' OD or Henley polyhedron, which is a triacontahedron with truncations along the fivefold directions (Henley, 1986). Along these shorter fivefold directions, the polyhedron directly connects to the B OD regions that generate Yb positions (see Fig. 3c in Takakura et al., 2007). Accordingly, for the present model a truncation of the sphere representing the hollow center was introduced: half spheres of radius $rB_1$, radially located at $rB_2 – rB_1$ on all fivefold directions are redefined: if a cut with physical space takes place in these regions, Yb atoms are generated, i.e. they are taken as a part of the second shell.

Furthermore, regions in the outer shell of B along the twofold perpendicular-space directions are subjected to a twofold physical-space shift (light blue in Fig. 2a). This is conducted as follows: half spheres of radius $(rB_3 – rB_2)/2$ centered on the tangent of the OD at $rB_3$, and additional spheres of radius $(rB_3 – rB_2)/4$ centered at a center distance of 0.75 $rB_3$, are defined in all twofold directions of an icosahedron in perpendicular space. Atom positions generated by a cut of physical space with these regions are subjected to a shift of $a_0\tau^{-5} \approx$ 0.7 Å along the corresponding twofold physical-space direction. This rather elaborate definition of the regions allows for a narrow selection along the twofold directions with the use of spherical objects only.

The OD on the V site is constructed as a central sphere plus two concentric spherical shells (Fig. 2b). The inner sphere V1 consists of a small hollow center. Both further shells, V2 and V3, generate Cd positions in physical space. The outer geometry of the V OD in the Takakura model is formed by a spherical arrangement of 32 Henley polyhedra (Fig. 3a in Takakura et al. 2007) and can be approximated by a single large sphere. Analogous to the case of the B OD, the outer shell V3 contains regions that are subjected to a physical-space shift (light blue in Fig. 2b). These regions, defined as half spheres of radius $rV_3 – rV_2$ centered on the tangent of the OD at $rV_3$, are located on all threefold directions of an icosahedron in perpendicular space. Atom positions generated by a cut of physical space with these regions are subjected to a shift of $a_0\tau^{-5}$ along the corresponding threefold physical-space directions.

The site symmetry of the E edge center is lower than that of the V and B sites, and the ideal polyhedral OD for this position in the 6D unit cell is a rhombic icosahedron (Yamamoto, 1996). As an approximation we use an ellipsoid, in equivalence to the spheres on the higher-symmetry positions. This ellipsoid is assumed to have two different radii with an aspect ratio corresponding to that of a rhombic icosahedron. The height of the latter is $h = a\sqrt{5}$, and the radius is $r_m = 2a\tau/\sqrt{5}$, where $a$ is the polygon edge length. This leads to an aspect ratio $h/r_m = 5/4\tau$. In the present model, the short axis, representing the height of the rhombic icosahedron, is handled as an adjustable parameter rE, while the long axes are held at $4\tau/5$rE. The short axis of each of the six E ODs is oriented along a fivefold direction in perpendicular space (Fig. 2c).

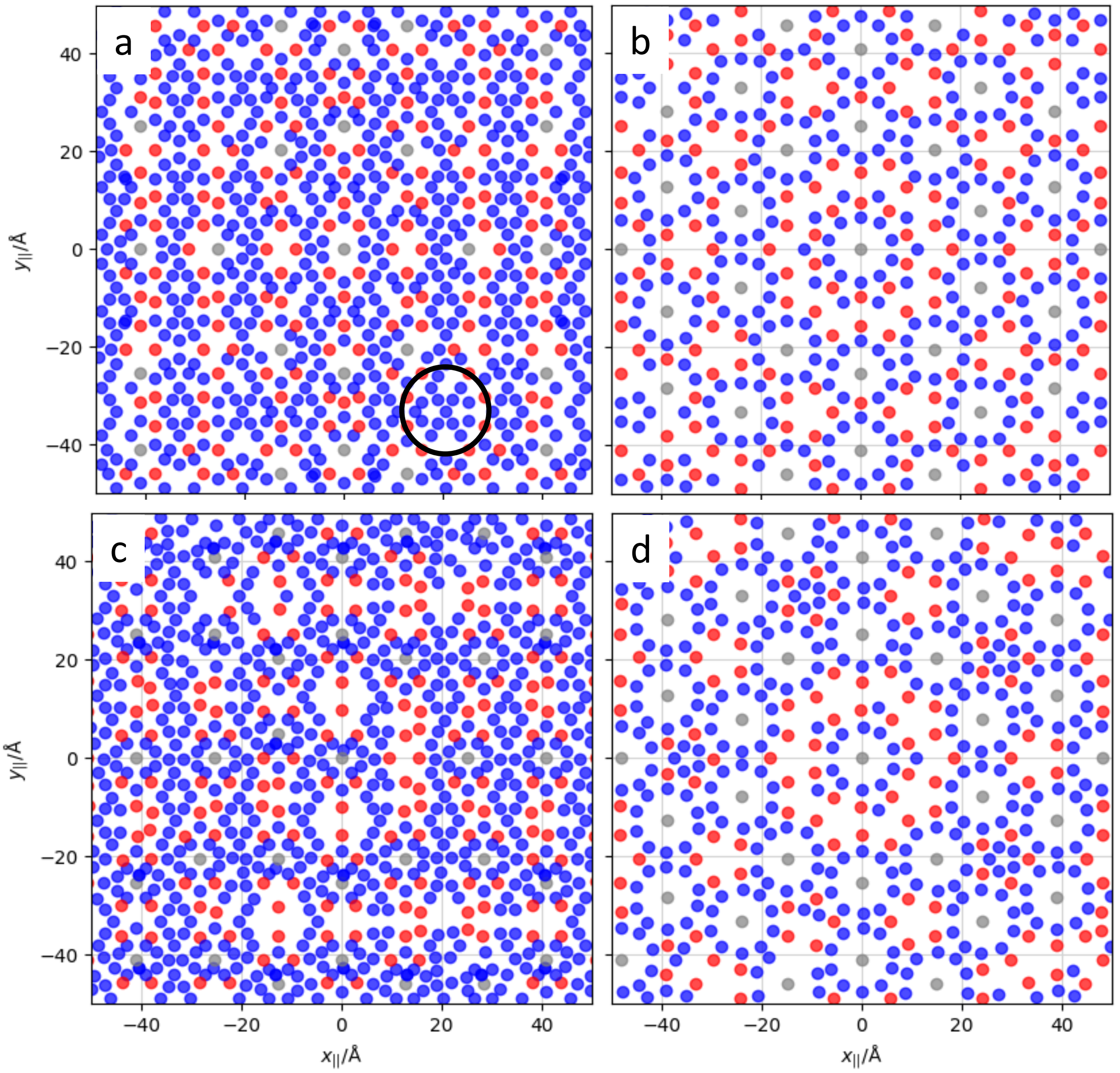

**Figure 3** Physical-space structure generated by the model in a twofold (a) and fivefold (b) plane at z = 0. For comparison: twofold- (c) and fivefold- (d) plane structures generated by the Takakura model.

Note that the use of an ellipsoidal E OD implies that the spherical truncations of the outer shell of the B OD represent an approximation: these truncations were introduced in order to avoid short distances generated by neighbored E and B ODs and hence should also be ellipsoidal. This, however, would lead to a significant increase in calculation time (see discussion). Since, on the other hand, spherical truncations lead to a very low occurrence of short distances and a density close to the experimental value, they were implemented in the model for the sake of calculation costs.

### 3. Results

Assuming the conditions provided in section 2 as fixed, the radii of the shells remain as the seven free parameters of the model. These can be varied such that the resulting physical-space structure adjusts to criteria defined by the user according to the physical problem at hand. The set of parameters provided in Table 1 was obtained by manual adjustment with the following demands: i) reproduction of the experimental chemical composition $Cd_{84}Yb_{16}$ and mass density of 8.88 g/cm$^3$ (Guo et al, 2000). ii) reproduction of the physical-space atom arrangement, in particular the reproduction of Tsai cluster shells and the distribution of Tsai-cluster centers as reported in Takakura et al. (2007) and De Boissieu et al. (2007) and the avoidance of unphysically short distances. iii) reproduction of the experimental electron density maps obtained by X-ray single-crystal diffraction (Takakura et al., 2002, Takakura et al., 2007).

**Table 1** Values obtained for the adjustable parameters of the spherical model, leading to a good fit to experimental and previous structure data.

| $rV_1/a_0$ | $rV_2/a_0$ | $rV_3/a_0$ | $rB_1/a_0$ | $rB_2/a_0$ | $rB_3/a_0$ | $rE/a_0$ |
|---|---|---|---|---|---|---|
| 0.1 | 0.879 | 1.143 | 0.335 | 0.940 | 1.490 | 0.492 |

Configured with the parameters in Table 1 the model assumes a composition of $Cd_{84.2}Yb_{15.8}$ and a density of 8.20 g/cm$^3$. These values represent the average of calculation runs for spherical atom arrangements with a diameter of 100 Å containing about 21.500 atom positions and cubic arrangements with an edge length of 100 Å containing about 40.500 atom positions, at different placements of the origin of the physical- and perpendicular-space coordinate system with respect to the 6D lattice.

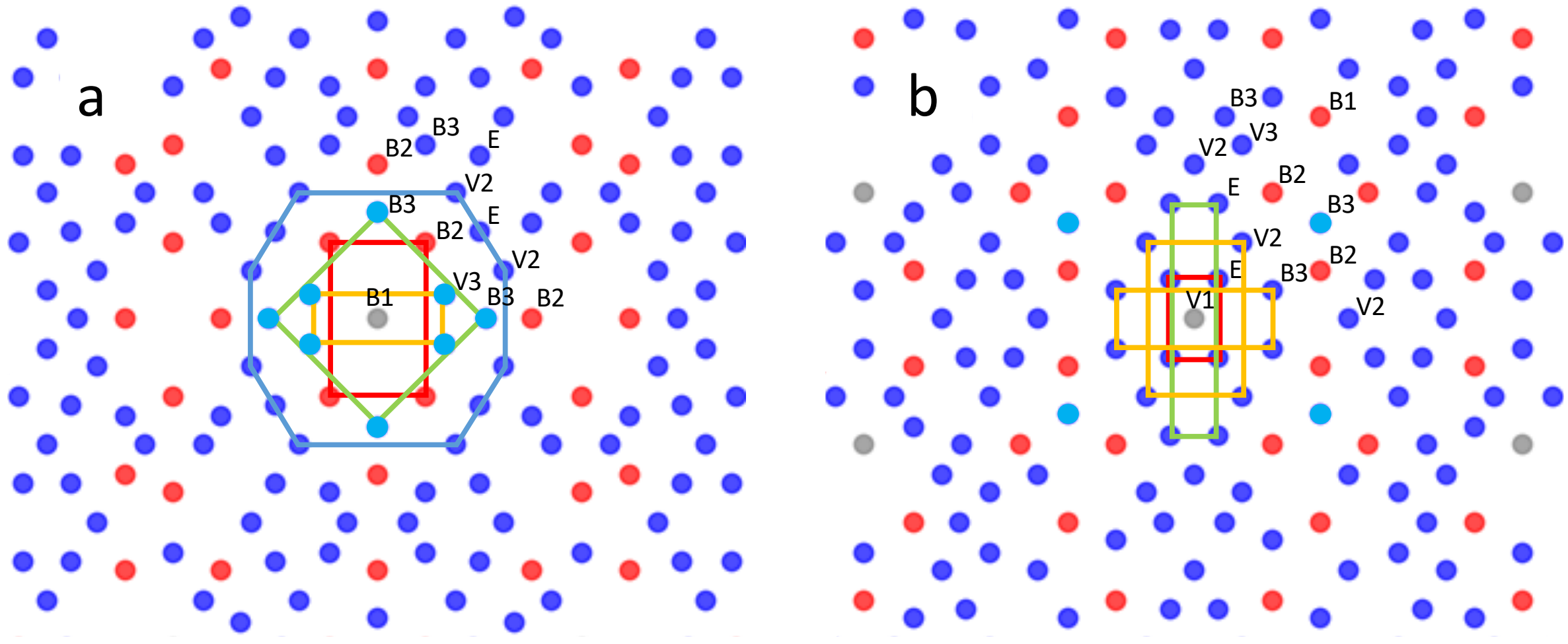

**Figure 4** Physical-space structure generated by the model with the origin of the physical- and perpendicular-space coordinate systems on the body-centred position (a) and the corner position (b) of the 6D lattice. Traces of a Tsai cluster and a Bergman cluster are indicated by polygons in (a) and (b) respectively. The labels denote the OD shell generating the position, Cd atoms subjected to a physical-space shift are displayed in lighter blue.

Fig. 3 shows the physical-space structure generated by the model in the twofold plane (a) and the fivefold plane (b) at z = 0. Cd atoms and Yb atoms are shown in blue and red, respectively. Grey markers stand for vacant positions created by the central spheres of the B and V ODs. For direct comparison, the corresponding results calculated with the model by Takakura et al. (2007) are shown in (c) and (d).

Fig. 4 shows local regions of the twofold plane at z = 0 calculated with the origin of the physical- and perpendicular-space coordinate system on the body centered ½(1 1 1 1 1 1) position (a) and the (0 0 0 0 0 0) corner position (b). The figures contain polygons corresponding to the traces of concentric clusters in the plane z = 0. In (a), the blue outermost octagon corresponds to the trace of a triacontahedron with Cd atoms on the vertices and edge-center positions. The green square connecting Cd atoms on the twofold axes corresponds to the trace of an icosidodecahedron, the red rectangle to that of an Yb icosahedron and the yellow rectangle to that of a Cd dodecahedron. These clusters constitute the shells of a Tsai cluster, the most prominent building block of the structure. The Tsai-cluster shells, as generated by the present model, are shown in Fig. 5a. It contains 154 atoms altogether (not taking into account the inner tetrahedron occurring in different orientations). Note that the size of the dodecahedron and the icosidodecahedron are directly set by the physical-space shifts along threefold and twofold directions of the V and B ODs, respectively. These clusters appear at a high density in the structure, and their centers are located at the grey positions in Fig. 3.

In Fig. 4a, in one quadrant of the cluster the atom positions are labelled according to their generating OD and shell. The cluster center, e.g. is generated by the (hollow) core of the B OD and is correspondingly labelled B1, the positions generated by the second shell are labelled B2, etc.. Cd positions subjected to a physical-space shift are displayed in lighter blue.

It is seen that the Cd dodecahedron and icosidodecahedron are generated by the outer shell of the V and B OD, respectively. The Yb icosahedron is generated by the second shell of the B OD (which is the only hyperspace element generating Yb positions), and the large Cd triacontahedron has vertex positions generated by the second shell of the V OD and edge-center positions generated by E ODs.

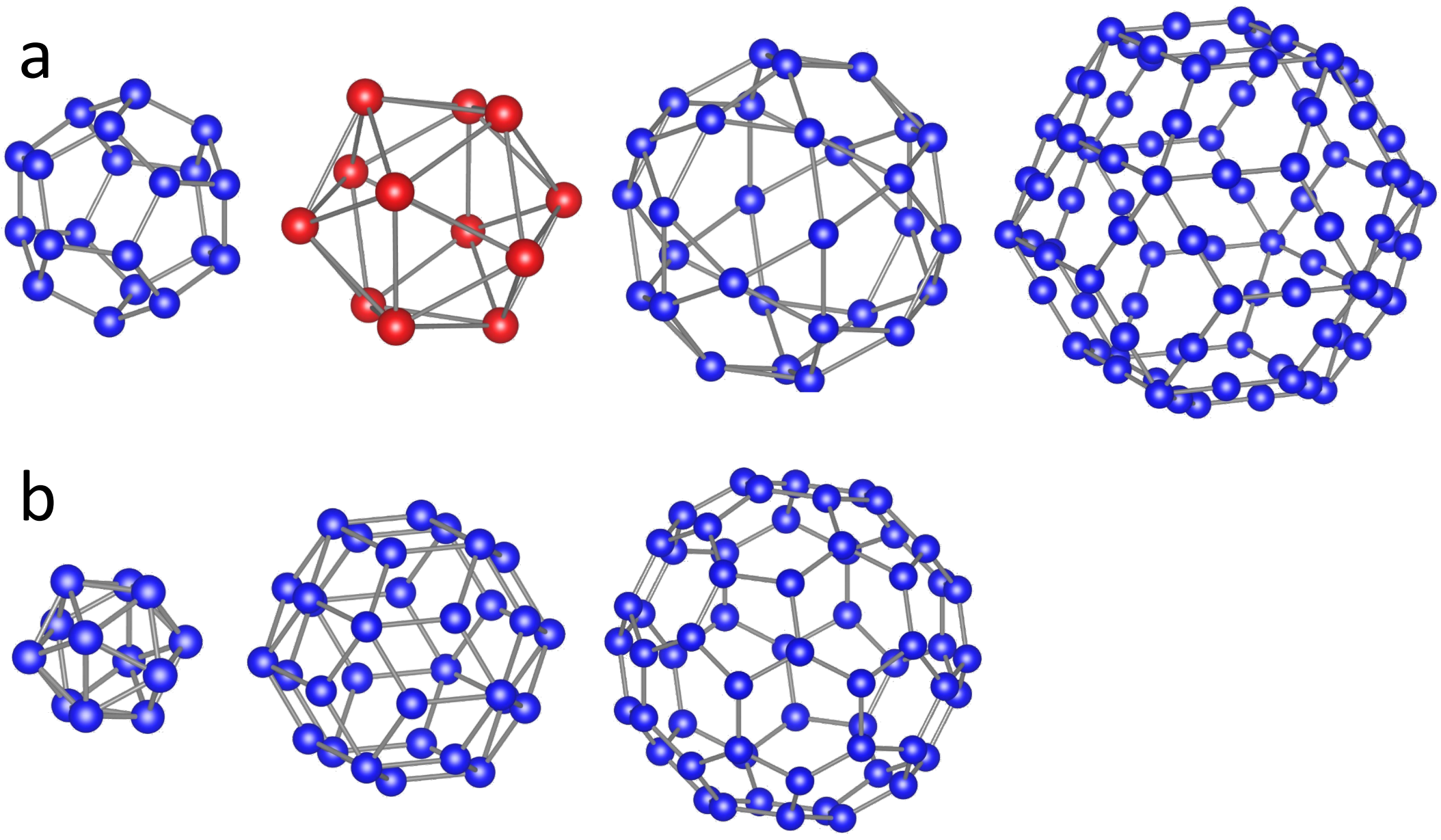

**Figure 5** (a) Tsai cluster as generated by the model, consisting of concentric shells forming a Cd dodecahedron, a Yb icosahedron, a Cd icosidodecahedron and a triacontahedron with Cd atoms on the vertices and edge-center positions (left to right). (b) Bergman cluster generated by the model, consisting of concentric shells forming a Cd dodecahedron, a triacontahedron with Cd atoms on the vertices and a Cd truncated icosahedron.

In Fig. 4 (b) the polygons depict traces of a truncated icosahedron (green), a triacontahedron, which is represented by two rectangles, emphasizing the fact that it can be decomposed into an icosahedron and a dodecahedron (yellow), and a small icosahedron (red). These clusters constitute the 104 atom Bergman cluster, which is shown in Fig. 5(b), a second structural element of i-Cd-Yb. Since the volume of the empty inner sphere of the V OD is very small, most instances of the Bergman cluster have a central Cd atom, only occasionally this position remains empty (labelled V1 in Fig. 4b). One instance of a Bergman cluster is indicated by a black circle in Fig. 3(a).

Fig. 4b reveals that the central icosahedron of the Bergman cluster is generated by E ODs. The dodecahedron inscribed in the triacontahedron is generated by the second shell of the V OD, V2, and the icosahedron inscribed in the triacontahedron by the third shell of the B OD, B3. The outer truncated icosahedron is generated by E ODs.

In Fig. 6, the structure generated by the model is compared with the 2D electron density as determined by X-ray diffraction by Takakura et al. (2002). The experimental electron density in a twofold plane is shown in the right half of Fig. 6. On the left, the structure as generated by the model is depicted. Most atom positions are reproduced correctly and at most high-intensity spots in the electron density correspond to the heavy Yb generated by the model.

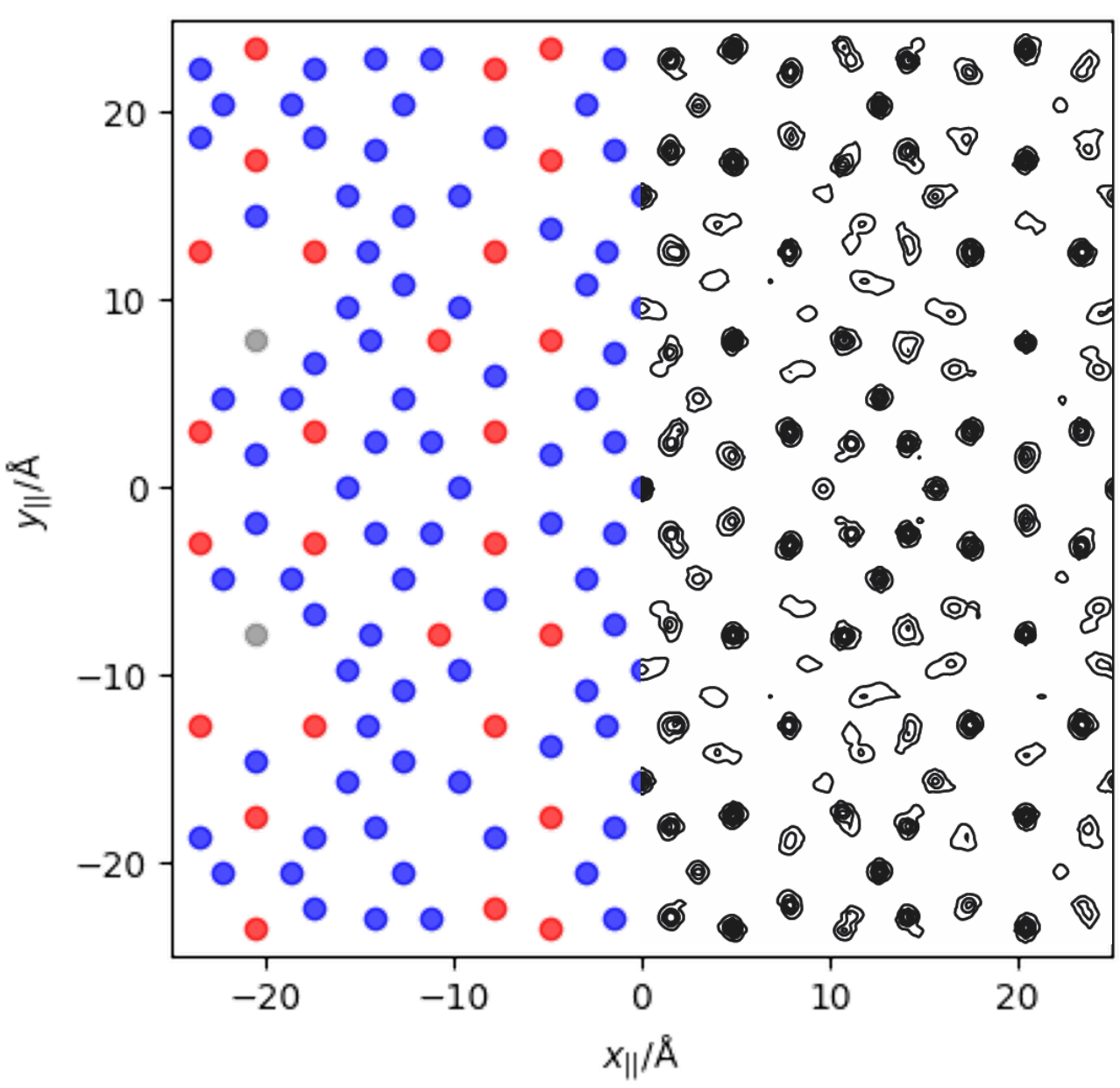

**Figure 6** Comparison of the physical-space structure in a twofold plane generated by the model (left half) with a corresponding 2D cut of the electron density measured by X-ray diffraction (right half, courtesy of H. Takakura).

The model does generate some extra Cd positions in this plane and a few positions show slight shifts. However, generally the agreement between the model structure and the experimental electron density is very good. A similar comparison of the generated physical-space structure shown in Fig. 4a with the 2D cut of the electron density measured by X-ray diffraction shown in Fig. 2b of Takakura et al. (2007) can be made. In this case the agreement is also excellent.

Some features of the perpendicular-space structure generated by the model are depicted in Fig. 7. Perpendicular-space coordinates of Cd atoms are shown in blue, those of Yb atoms are shown in red. The scale bar implies an identical scaling in physical and perpendicular space. Fig 7a, from left to right, displays central cuts of the V, B, and two instances of E ODs at $z_{\perp}= 0$. The V and B ODs are shown along twofold, the E ODs along fivefold directions in perpendicular space. Circles mark the radii of the shells of the OD, for E circles of the large ellipsoidal radius are shown. For the V and B ODs, the positions which are associated to a physical-space shift of the Cd atoms are shown in lighter blue, they are located in the threefold and twofold regions of the outer shell of V and B, respectively. The empty regions in the outer shell of the B OD correspond to the truncation along the fivefold directions. The two E ODs displayed assume different orientation with respect to the direction of view. The left instance is shown from a direction making an angle (of 26.5 °) with the short radius, so its ellipsoidal shape can be seen. The right instance is seen along the short radius, hence it appears round.

Fig. 7 a can be directly compared to Fig. 2. By construction, all perpendicular-space positions generated by the model are located within the bounds defined by the ODs.

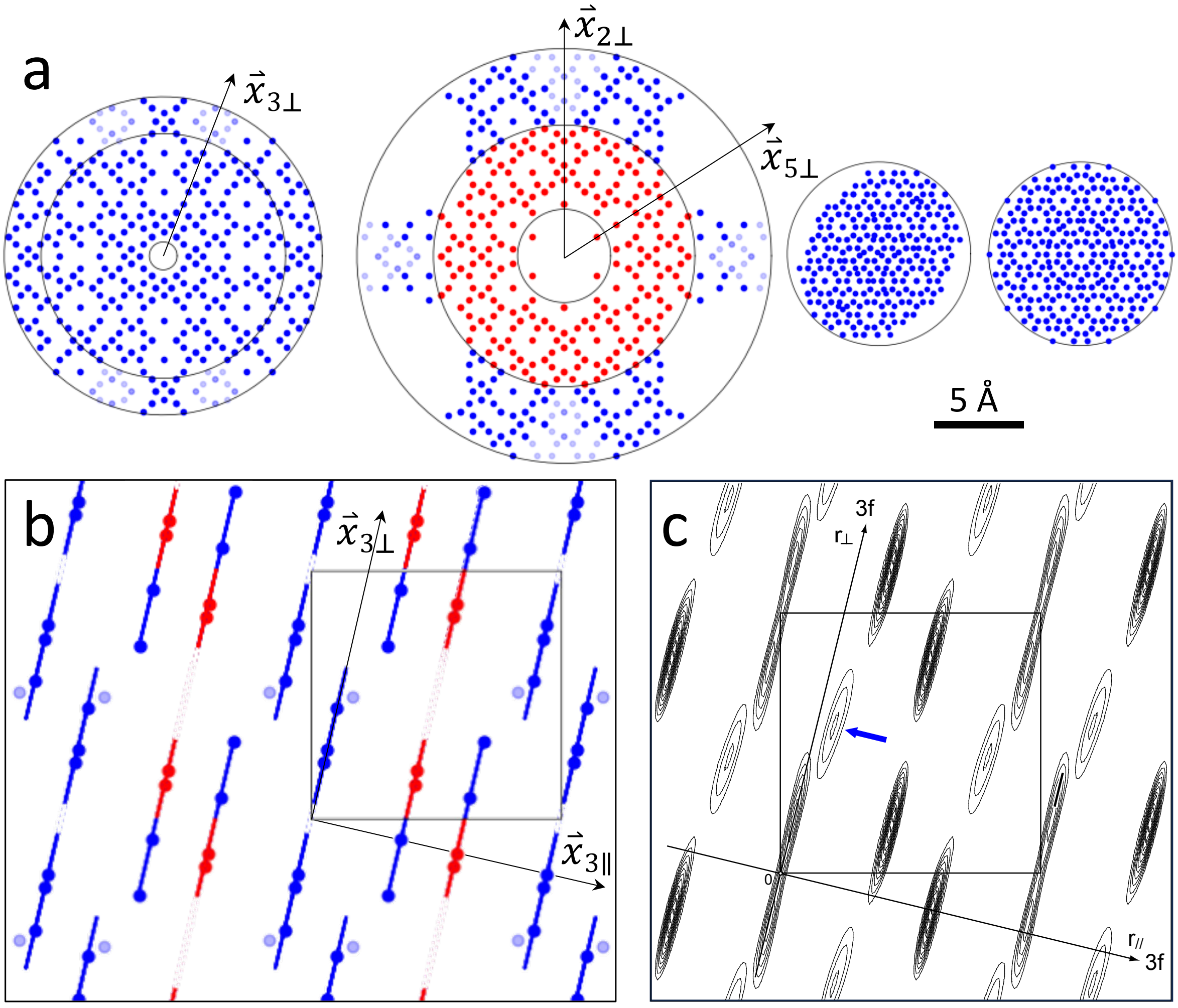

**Figure 7** Perpendicular-space structure generated by the model. Blue: Cd; Red: Yb; perpendicular-space-shifted positions shown in lighter blue. (a) Left to right: central cuts of the V, B along twofold, and of two instances of E ODs along fivefold directions in perpendicular space. (b) Threefold rational section displaying the coordinates generated by the model (points) and the traces of the spherical ODs (lines). (c) Experimental electron-density map in a threefold rational section (courtesy of M. De Boissieu and H. Takakura).

Fig. 7b depicts the generated perpendicular-space coordinates in a threefold rational section as points and, as a guide to the eye, the traces of the basic V and B ODs are as lines (c.f. Fig 1c). In this representation, the physical-space shift the outermost points generated by the V ODs are subjected to, can be seen. These points, which are shown in lighter blue are shifted along the threefold physical-space direction and hence are displayed outside of the traces representing the basic spherical V ODs. For direct comparison, Fig. 7c displays the experimental electron-density map in a threefold rational section (De Boissieu et al., 2007), which displays shifts of the exterior of the V ODs (arrow). It can clearly be seen, that the shifted atom positions generated by the model perfectly fall into the shifted regions in the experimental electron-density map.

#### 4. Discussion

With the set of parameters in Table 1, the mass density calculated by the model is 0.6 g/cm$^3$ lower than the experimental density, i.e. off by less than 7.0 %. The chemical composition deviates by 0.2 at.% from the experimental values (Guo et al., 2000). These characteristics are thus very accurately reproduced. Comparison of the physical-space structure generated by the model with experimental values of the electron density obtained by X-ray diffraction also shows highly satisfactory agreement. Moreover, the cluster substructure of i-CdYb is very well reproduced – the model generates Tsai clusters, the main building blocks of the structure, with the correct concentric shells at the correct positions. Furthermore, Bergman clusters with their characteristic concentric shells are generated. Bergman clusters can also serve as building blocks in an alternative structural description (Buganski et al., 2024).

Critical comparison with the Takakura model reveals that the spherical model generates some difference mainly in the Cd positions. Additional Cd atoms are generated in the fivefold plane (Fig. 3(d)) along the pseudo-twofold directions axes around the central ring of Yb atoms. While the refined structure has only 5 Cd atoms here, the present model generates 10. Also, the Takakura model has five small pentagons in five of the Yb pentagons at a radius of about 30 Å from the center, whereas in the present model, these Yb pentagons are all empty. The positions on the inscribed dodecahedron of the Bergman cluster (yellow rectangle in Fig. 4b) labelled B3, are occupied by Cd atoms in the present model configuration. In the Takakura model these positions are occupied by Yb, and together with the neighbored Yb atoms labelled B2, constitute the Yb pair in the truncated Cd tetrahedra of the AR unit (see below). The main characteristics of the refined structure, however, are reproduced by the model, including the positions of the Tsai-cluster centers shown in grey. The major structural characteristics of the twofold plane (Fig. 3(c)) of the refined structure are well reproduced as well, including the positions of the Tsai-cluster centers. The largest difference stems from the fact that the present model does not reproduce the central Cd tetrahedron of the Tsai clusters. As a result, for each Tsai cluster, up to four Cd atoms are missing. In the Takakura model, the inner tetrahedron occurs in different orientations, which cannot be handled by the present model.

This real-space comparison shows that the model reproduces the substantial structural features remarkably well, but there are some discrepancies with respect to the more accurate Takakura model. In order to quantify the degree of fidelity and reliability of the model, i.e. to promote it to a mature structure model, further experimental validation, in particular non-local comparison with diffraction data is required.

Analysis of the atomic distances show peaks around 3.52 and 5.98 Å and the shortest distances at about 2.17 Å. Depending on the placement of the origin of the physical- and perpendicular-space coordinate system, the model generates a very low number of unphysically short distances of 0.67 Å. The frequency which these short distances occur correspond to a density of excess atoms in the one-percent range. These atoms involved are generated by the outer peripheries of the B and the E OD, and occur at residual densities since the radius of the cut-out half sphere in the outer shell of B is slightly smaller than the long ellipsoidal radius of OD E. It should be noted, that 0.67 Å short distances also occur in the experimental electron density map (right-hand side of Fig. 6). The elongated spots in the map correspond to split positions of this exact distance. These can be included in the model by decreasing the size of the cutouts along the fivefold directions in the B ODs.

Note that the described structural features are a direct consequence of the particular set of parameters in Table 1, which result from the demands described in section 3. Depending on the individual application, other configurations of the model may be more appropriate. For example, the physical-space structure of the Takakura model, besides the Tsai clusters (referred to as RTH therein), contains further building units referred to as AR which define pairs of Yb atoms. The AR units can combine to form larger polyhedra containing atom positions that are not within the Tsai clusters. One of these is a stellate polyhedron with a central Bergman cluster. In the Takakura model, the dodecahedron inscribed in the Bergman clusters' triacontahedral shell is occupied by Yb atoms. In contrast, the spherical model in its present configuration generates Cd atoms on these positions. If required, an occupation by Yb can be enforced by setting $rB_2 > 1.042\ a_0$, but this choice of parameters leads to a composition of about $Cd_{80}Yb_{20}$ , i.e. deviating further from the experimental value, and an almost quadrupled density of unphysically short distances.

If required, the spherical model can also be adapted so that it generates atom positions on the inner Tsai-cluster tetrahedron, which are not included using the set of parameters in Table 1, by changing $rV_3$ to $1.2\ a_0 - 1.5\ a_0$. This configuration, however, leads to the formation of a complete icosahedron with 20 occupied Cd positions comprising all possible orientations of the Cd tetrahedron at once. Accordingly, the actual Cd density in the cluster centers will be overestimated and unphysically short distances will occur.

These examples show that the simplicity of the spherical model comes with a cost: it can versatilely be adapted to the specific scientific problem at hand, but this always involves unwanted impairment of other properties of the structure. This is a significant difference to the Takakura model, which allows for covering all possible situations with a single set of parameters, without the necessity to make any compromises.

The main calculation is programmed in about 100 essential lines of non-optimized python code. Additionally the code contains numerous read-only lines containing definitions, coordinates, etc. A streamlined version of the code, containing the routines calculating the structure and generating a cif file as output is available in a Jupyter-notebook format (Feuerbacher, 2025). Calculation of a spherical arrangement of atoms with a radius of 50 Å containing 2753 positions takes about 12 seconds on a 2022 MacBook Pro Laptop with an M2 processor and 16 Gb of memory. A cube of 100 Å edge length, containing more than 40.000 positions takes about 540 seconds. The calculation time is dominated by the presence of the ellipsoidal E ODs. Without the E ODs, the calculation of the 50 Å sphere and the 100 Å atom cube take about 1.5 and 60 seconds, respectively. The implemented cut-detection routine, due to the orientation-dependence of the non-spherical E ODs, requires the solution of three scalar products and one cross product. These have to be calculated for each of the $(2n)^6$ 6D lattice nodes for each of the 6 E ODs. While for the small sphere of atoms a 6D lattice extension of n = 3 suffices, the large atom cube requires values of 6.

This observation retrospectively emphasizes the benefits of using mainly spherical features in the present calculation. When polyhedral ODs are used, typically two scalar products per limiting face of each polygon have to be solved. Even if symmetries are taken advantage of, i.e. only the polygons in the asymmetric units are calculated, polyhedral ODs require disproportionately longer calculation times. This argument may not hold for professional coders, being able to employ tricks to accelerate and optimize code, but most likely is relevant for the average materials scientist in need of a sheer working model.

In conclusion, in this paper I present a compact structure model for the icosahedral quasicrystal $Cd_{5.7}Yb$. It is based on a spherical approximation and is less accurate and less universal than the established Takakura model. It requires adaption to the specific scientific problem to be applied to, and cannot handle all potential situations with a single set of model parameters. Nevertheless, it can serve as alternative to the Takakura model for situations in which the knowledge of ultimate structural detail is less important than versatility, fast calculation and simple adaptability to the actual problem. The presented model can surely be further optimized, and it can serve as a basis for the development of models for related phases like icosahedral InAgYb and CdCa and other structures based on Tsai clusters.

**Acknowledgements** The author is greatly thankful for inspiring discussions with Hiroyuki Takakura, Tsunetomo Yamada, Marc De Boissieu, and Jianbo Wang.

**Conflicts of interest** The author declares that there are no conflicts of interest.

**Data availability** The program code in python containing the routines for calculating the structure and generating a cif file as output is available in a Jupyter-notebook format (Feuerbacher, 2025).

## References

Boudard, M., de Boissieu, M., Janot, C., Heger, G., Beeli, C., Nissen, H.-U., Vincent, H., Ibberson, R. M., Audier, M. & Dubois, J. M. (1992). *J. Phys.: Condens. Matter*, **4**, 10149 – 10168.

Bugański, I., Strzałka, R. & Wolny, J. (2024). *Acta Cryst. B*, **80**, 84 – 93.

Cahn, J.W., Shechtman, D. & Gratias, D. (1986). *J. Mater. Res.*, **1**, 13 – 26.

Cordelli, A. & Gallo, P. (1995). *J. Phys.: Condens. Matter*, **7**, 1245 – 1254.

De Boissieu, M., Janot, C. & Dubois, J.-M. (1990). *J. Phys.: Condens. Matter*, **2**, 2499 – 2517.

De Boissieu, M., Janot, C., Dubois, J.-M. & Dubost, B. (1991). *J. Phys.: Condens. Matter*, **3**, 1 – 25.

De Boissieu, M., Takakura, H., Pay Gómez, C., Yamamoto, A. & Tsai, A. P. (2007). *Philos. Mag.*, **87**, 2613 – 2633.

Duneau, M. & Katz, A. (1985). *Phys. Rev. Lett.*, **54**, 2688 – 2691.

Feuerbacher, M. (2025). *Zenodo*. https://doi.org/10.5281/zenodo.16942398

Guo, J. Q., Abe, E. & Tsai, A. P. (2000). *Phys. Rev. B*, **62**, R14605 – R14608.

Henley, C. L. (1986). *Phys. Rev. B*, **34**, 797 – 816.

Ishimasa, T., Oyamada, K., Arichika, Y., Nishibori, E., Takata, M., Sakata, M. & Kato, K. (2004). *J. Non-Cryst. Solids*, **334–335**, 167 – 172.

Kramer, P. & Neri, R. (1984). *Acta Cryst.* A, **40**, 580 – 587.

Ledieu, J., McGrath, R., Diehl, R. D., Lograsso, T. A., Delaney, D. W., Papadopolos, Z. & Kasner, G. (2001). *Surface Science*, **492**, L729 – L734.

Levine, D., Lubensky, T. C., Ostlund, S., Ramaswamy, S., Steinhardt, P. J. & Toner, J. (1985). *Phys. Rev. Lett.*, **54**, 1520 – 1523.

Quinquandon, M. & Gratias, D. (2006). *Phys. Rev. B*, **74**, 214205.

Shechtman, D., Blech, I., Gratias, D. & Cahn, J. W. (1984). *Phys. Rev. Lett.*, **53**, 1951 – 1953.

Takakura, H., Yamamoto, A., Shiono, M., Sato, T. J. & Tsai, A. P. (2002). *J. Alloys Compd.*, **342**, 72 – 76.

Takakura, H., C.P. Gómez, Yamamoto, A., De Boissieu, M. & Tsai, A. P. (2007). *Nat. Mater.*, **6**, 58 – 63.

Takakura, H., private communication. (2025).

Uryu, H., Yamada, T., Kitahara, K., Singh, A., Iwasaki, Y., Kimura, K., Hiroki, K., Miyao, N., Ishikawa, A., Tamura, R., Ohhashi, S., Liu, C. & Yoshida, R. (2024). *Adv. Sci.*, **11**, 2304546.

Yamamoto, A. (1996). *Acta Cryst.* A**52**, 509 – 560.

Yamamoto, A., Takakura, H. & Tsai, A. P. (2003). *Phys. Rev. B*, **68**, 094201.

Yang, W., Feuerbacher, M., Tamura, N., Ding, D., Wang, R. & Urban, K. (1998). *Philos. Mag. A*, **77**, 1481 – 1504.